\documentclass[conference]{IEEEtran}

\usepackage[T1]{fontenc}
\usepackage[utf8]{inputenc}
\usepackage{stmaryrd} %
\usepackage{etoolbox}
\usepackage[colorlinks,urlcolor=blue,linkcolor=black,citecolor=black]{hyperref}

\patchcmd{\thebibliography}{\footnotesize}{\fontsize{7}{7.6}\selectfont}{}{\errmessage{bib font patch failed}}
\patchcmd{\thebibliography}{\vskip 0.3\baselineskip plus 0.1\baselineskip minus 0.1\baselineskip}{\vskip 2pt}{}{\errmessage{bib skip patch failed}}

\begin{document}

\title{QECirc: A Community-Driven, FAIR Library of\\ Quantum Error Correction Circuits\\[2pt]
  \large\normalfont on behalf of the QECirc contributors}

\author{%
  \IEEEauthorblockN{Ludwig Schmid\textsuperscript{1}, Tom Peham\textsuperscript{1}, and Robert Wille\textsuperscript{1,2}}
  \IEEEauthorblockA{\textsuperscript{1}Chair for Design Automation, Technical University of Munich, Munich, Germany\\
    \textsuperscript{2}MQSC, Garching near Munich, Germany \\
    contribute@qecirc.com \quad \href{https://qecirc.com}{qecirc.com}}%
}

\maketitle

\begin{abstract}
Quantum error correction (QEC) underpins fault-tolerant quantum computing, but the circuits that realize QEC codes (encoders, logical state-preparation and gate routines, and syndrome-extraction schedules) are scattered across papers and one-off repositories in incompatible formats, often documented too sparsely to reconstruct or reuse. We present \emph{QECirc}, a community-driven, open library that makes QEC circuits Findable, Accessible, Interoperable, and Reusable (FAIR). Each entry links to its underlying code, ships in Stim and OpenQASM with interactive visualizations, and is checked for correctness before publication. In short, where a catalog such as the Error Correction Zoo describes the codes, QECirc provides the concrete circuits that realize them.
\end{abstract}

\section{Introduction and Motivation}
Useful quantum computing is widely expected to require quantum error correction (QEC), in which logical qubits are redundantly encoded into entangled states of many noisy physical qubits~\cite{AF:22}. Realizing this requires more than a good code: it requires concrete \emph{circuits} that prepare encoded states, map data into and out of the code space, extract error syndromes, and apply logical gates. Preparing logical states is a central primitive for universal fault-tolerant computation, and its cost, especially its two-qubit-gate count and depth, contributes directly to the overhead of workflows such as magic-state distillation and cultivation~\cite{PSWH:26}.

Consequently, encoder and logical state-preparation synthesis is an active research area: reinforcement learning for hardware-adapted fault-tolerant circuits~\cite{ZOC+:25}, greedy, rollout-based, and SMT-based encoder synthesis~\cite{PSWH:26}, and AI-optimized graph decimation~\cite{DPB+:26}. These works report substantial reductions in gate count and depth for codes from the Steane and Golay codes~\cite{Ste:96} to the 144-qubit gross code~\cite{BCG+:24}, yet the resulting circuits are typically buried in paper appendices, ad-hoc scripts, or private repositories, in whatever format a particular group happens to use. As a result, researchers repeatedly re-derive the same circuits, and cross-work comparisons remain difficult. Code catalogs such as the Error Correction Zoo~\cite{AF:22} have become standard references for QEC \emph{codes}, but no analogous format-agnostic resource exists for the \emph{circuits} that implement them.

QECirc fills this gap. It is a community-driven, open library of QEC circuits, browsable at \href{https://qecirc.com}{qecirc.com}, designed around the FAIR principles for scientific data~\cite{WDA+:16}, in which every published circuit is checked for correctness and every code carries its stabilizer check matrices for direct reproducibility.

\section{Background and Related Work}
Stim~\cite{Gid:21} is a de-facto standard for stabilizer circuits and their fast simulation, OpenQASM~\cite{CJA+:22} is widely used for expressing quantum circuits, and tools such as Quirk and Crumble~\cite{Gid:qk} help inspect small circuits interactively. The Error Correction Zoo~\cite{AF:22} catalogs codes, and toolkits such as the Munich Quantum Toolkit (MQT)~\cite{WBF+:24} construct and decode them.

The circuits themselves are the focus of an active research community. A wide range of methods synthesizes and optimizes encoders and logical state-preparation circuits~\cite{PSWH:26, ZOC+:25, DPB+:26, PSB+:25, RCKP:20, SPB+:25, CZO+:26, FA:25}, fault-tolerant syndrome-extraction circuits and gadgets~\cite{LPZ+:26, CR:18, CB:18, KLY+:25}, and logical gates from code automorphisms~\cite{SKW+:24}, building on classical and learned Clifford and linear-reversible (CNOT) synthesis routines~\cite{CLR+:26, BM:21, PMH:08, WKB:25, GMV:25, GM:24}. QECirc is \emph{complementary}: rather than reproducing a taxonomy or synthesis algorithm, it collects, standardizes, and republishes the circuits such methods produce on one searchable platform.

\section{The QECirc Library}
\textbf{Data model.} QECirc links four entities: \emph{codes}, \emph{circuits}, the \emph{tools} that generate them, and the \emph{papers} that describe them. A code is identified by its parameters $\llbracket n,k,d\rrbracket$, descriptive tags, and its stabilizer and logical check matrices. Codes are canonicalized on entry, so different descriptions of the same code collapse to a single entry. Each circuit is attached to a code and a task: encoding, logical Pauli-eigenstate preparation, fault-tolerant flag gadgets, or logical Clifford gates. Every circuit records its provenance (the generating tool and originating paper), computed metrics such as qubit count, depth, and two-qubit gate count, plus connectivity and device tags, and reminds the user to cite the original work on copy or download.

\textbf{Tools.} A curated directory catalogs the software used to produce the circuits, ranging from reinforcement-learning circuit discovery~\cite{ZOC+:25}, fault-tolerant state-preparation and encoding synthesis~\cite{PSB+:25, SPB+:25, PSWH:26}, and Clifford and CNOT synthesis engines~\cite{WKB:25, GMV:25, GM:24} to flag gadgets~\cite{FA:25}, automorphism-based logical gates~\cite{SKW+:24}, and qLDPC toolkits~\cite{KLY+:25}. Each tool is linked to its repository, its papers, and the circuits it contributed, crediting the generating methods.

\textbf{Interface.} Circuits are presented grouped by code. A ranked search finds entries by code name, alias, tag, or source paper; codes can be filtered by parameters and tags, circuits by their metrics, sorted, or flagged as favorites. Circuit views preserve qubit coordinates and can overlay derived detectors and logical observables. Each circuit can be exported to Stim or OpenQASM and links to live Crumble and Quirk views for in-browser reuse.

\section{Verification and Reproducibility}
Every submission is processed by an automated ingestion pipeline that extracts metrics, identifies the implemented code, and validates correctness before publication. It checks that encoders map the all-zero input into the code space and that state preparations satisfy every stabilizer and prepare the claimed logical basis state, for CSS and non-CSS codes alike, and derives detector and observable annotations for Stim-based simulation. Since each code exposes its check matrices, users can independently verify any result.

\section{FAIR and Community}
These properties make QECirc a FAIR~\cite{WDA+:16} resource for QEC circuits. Entries are \emph{findable} by code parameters, task, and tags, each with a stable identifier and machine-readable metadata for search engines and AI assistants; \emph{accessible} through an open website and public GitHub repository, with no login or paywall; \emph{interoperable} through standard formats (Stim and OpenQASM) that common simulators consume; and \emph{reusable}: openly licensed, with provenance and tool attribution on every entry.

QECirc is moreover developed as open community infrastructure. Contributions are submitted through guided templates and validated automatically, and the platform is open source, so the community can extend both the collection and the site. The first external contributions~\cite{CZO+:26} have already been merged and credited on the site. Such a shared resource reduces duplicated effort, enables fair comparison of synthesis methods against a common baseline, and lowers the barrier to obtaining a working circuit for a given code and task.

\section{Outlook: A Zoo for Circuits}
Our longer-term aim is for QECirc to become, for QEC \emph{circuits}, what the Error Correction Zoo~\cite{AF:22} is for codes: a single, trusted, community-driven reference where well-engineered circuits are found, shared, and improved. Planned work includes broader task coverage (syndrome-extraction schedules~\cite{LPZ+:26}), larger qLDPC code families such as bivariate-bicycle codes~\cite{BCG+:24}, richer hardware-aware metadata, and deeper cross-linking with the Zoo. QECirc is in active development, already spanning hundreds of verified circuits across a wide range of codes and tasks, and we invite the wider quantum error correction community to browse, use, and, above all, contribute at \href{https://qecirc.com}{qecirc.com}.

\section*{Acknowledgment}
{\scriptsize
We thank the QECirc contributors (everyone in the QEC community who has submitted, reviewed, or improved an entry), whose collective effort makes this library possible; the current list is maintained at \href{https://qecirc.com}{qecirc.com}. QECirc is supported by a Unitary Foundation micro-grant (\url{https://unitary.foundation/grants/2026_qecirc}) and by the Chair for Design Automation at the Technical University of Munich. The authors acknowledge funding from the European Research Council (ERC) under the European Union's Horizon 2020 research and innovation program grant agreement No.~101001318 and No.~101114305, and the Munich Quantum Valley (MQV), which is supported by the Bavarian state government with funds from the Hightech Agenda Bayern Plus. Furthermore, this work was supported by the BMFTR under grant numbers 13N17298 and 01MQ25001I, the Deutsche Forschungsgemeinschaft (DFG, German Research Foundation) under grant numbers 563402549 and 563436708. \emph{GenAI:} an initial draft of this abstract was prepared with a generative AI system (Anthropic Claude) and then reworked and finalized by the authors, who take responsibility for all content.\par
}

\bibliographystyle{IEEEtran}
\bibliography{references}

\end{document}